\documentclass{elsart}

\usepackage{amsmath}
\usepackage{graphicx}
\usepackage{ulem}             
\usepackage{color}

\def\etal{{\it et~al. }}
\def\J2{$J_{2\odot}$}
\def\cmr2{C/MR$^{2}$}
\def\d22{$C_{22}$}
\def\c20{$C_{20}$}
\def\32{$3$:$2$}

\begin{document}

\begin{frontmatter}



\title{On the oscillations in Mercury's obliquity}


\author[label1]{E. Bois}
\ead{Eric.Bois@oca.eu}
\author{and}
\author[label2,label3]{N. Rambaux} 
\ead{Nicolas.Rambaux@oma.be}

\address[label1]{Observatoire de la C\^ote d'Azur, UMR CNRS Cassiop\'ee, B.P. 4229, F-06304 Nice Cedex 4, France}
\address[label2]{Royal Observatory of Belgium, 3 Avenue Circulaire, B-1180 Brussels, Belgium}
\address[label3]{Depart. of Mathematics, Facult\'es Univ. N.D. de la Paix, 8 Rempart de la Vierge, B-5000 Namur, Belgium}

\vspace{2cm}
\centerline{Email address : Eric.Bois@oca.eu}
\centerline{Email address : Nicolas.Rambaux@oma.be}
\vspace{2cm}
\centerline{Telephone Number : + 33 [0]4 92 00 30 20 (E. Bois)}
\centerline{Telephone Number : +32 [0]2 373 67 33 (N. Rambaux)}
\vspace{2cm}
\centerline{Number of manuscript pages: 30}
\centerline{Number of manuscript tables: 3}     
\centerline{Number of manuscript figures: 8}

\newpage    
\vspace{5cm}
\centerline{Eric Bois}
\centerline{Observatoire de la C\^ote d'Azur,}
\centerline{UMR CNRS Cassiop\'ee,}
\centerline{ B.P. 4229, F-06304 Nice Cedex 4, France}
\vspace{0.5cm}
\centerline{Telephone Number : + 33 [0]4 92 00 30 20}
\centerline{Fax Number : + 33 [0]4 92 00 31 18}
\centerline{Email address : Eric.Bois@oca.eu}
\vspace{1cm}
\centerline{Running title: Librations in Mercury's obliquity}

\newpage
\begin{abstract}
Mercury's capture into the 3:2 spin-orbit resonance can be explained as a result of its chaotic orbital dynamics. One major objective of MESSENGER and BepiColombo spatial missions is to accurately measure Mercury's rotation and its obliquity in order to obtain constraints on internal structure of the planet. Analytical approaches at the first order level using the Cassini state assumptions give the obliquity constant or quasi-constant. Which is the obliquity's dynamical behavior deriving from a complete spin-orbit motion 
of Mercury simultaneously integrated with planetary interactions?
We have used our SONYR model (acronym of Spin-Orbit $N$-bodY Relativistic model) integrating the spin-orbit $N$-body problem applied to the solar System (Sun and planets). 
For lack of current accurate observations or ephemerides of Mercury's rotation, and therefore for lack of valid initial conditions for a numerical integration, we have built an original method for finding the libration center of the spin-orbit system and, as a consequence, for avoiding arbitrary amplitudes in librations of the spin-orbit motion as well as in Mercury's obliquity. 
The method has been carried out in two cases: (1) the spin-orbit motion of Mercury in the 2-body problem case (Sun-Mercury) where an uniform precession of the Keplerian orbital plane is kinematically added at a fixed inclination ($S_{2K}$ case), (2) the spin-orbit motion of Mercury in the $N$-body problem case (Sun and planets) ($S_{n}$ case). 

We find that the remaining amplitude of the oscillations in the $S_n$ case is one order 
of magnitude larger than in the $S_{2K}$ case, namely 4 versus 0.4 arcseconds (peak-to-peak). The mean obliquity is also larger, namely 1.98 versus 1.80 arcminutes, for a difference of 10.8 arcseconds. These theoretical results are in a good agreement with recent radar observations but it is not excluded that it should be possible to push farther the convergence process by drawing nearer still more precisely to the libration center. 
We note that the dynamically driven spin precession, which occurs when the planetary interactions are included, is more complex than the purely kinematic case. Nevertheless, in such a $N$-body problem, we find that the 3:2 spin-orbit resonance is really combined to a synchronism where the spin and orbit poles on average precess at the same rate while the orbit inclination and the spin axis orientation on average decrease at the same rate.
As a consequence and whether it would turn out that there exists an irreducible minimum of the oscillation amplitude, quasi-periodic oscillations found in Mercury's obliquity should be to geome\-trically understood as librations related to these synchronisms that both follow a Cassini state.
Whatever the open question on the minimal amplitude in the obliquity's oscillations 
and in spite of the planetary interactions indirectly acting by the solar torque on 
Mercury's rotation, Mercury remains therefore in a stable equilibrium state that 
proceeds from a 2-body Cassini state. 
\end{abstract}

\end{frontmatter}

\newpage

\section{Introduction} 

From radar measurements of the rotation of Mercury obtained by Pettengill $\&$ Dyce (1965), Colombo (1965) showed the non-synchronous rotation of Mercury and its 3:2 spin-orbit state (the rotational and orbital periods are 56.646 and 87.969 days respectively). 
The stability of the 3:2 spin-orbit resonance depends on a significant orbital eccentricity and the fact that the averaged tidal torque trying to slow the planet further is considerably less than the maximum averaged torque acting on the permanent axial asymmetry that causes the planet spin to librate about the resonant 3:2 value.
Moreover, Correia $\&$ Laskar (2004) have shown more recently that Mercury's 
capture into the 3:2 reso\-nance can be explained as a result of its chaotic orbital dynamics.

The upcoming missions, MESSENGER (Solomon \etal 2001) and Bepi\-Co\-lombo, with onboard instrumentations capable of measuring the rotational parameters of Mercury (see e.g. Milani \etal 2001), stimulate the objective to achieve a very accurate theory of Mercury's rotation. One aim of these missions is to determine the existence or not of a liquid core by measuring accurately both the rotation and the gravity field of Mercury. The current method for obtaining constraints on the state and structure of Mercury's core is based on the assumptions introduced by Peale in 1972 and revisited in 2002 (Peale \etal 2002). By measuring the amplitude of the 88-day libration in longitude, called $\varphi$ hereafter, the obliquity $\bar{\eta}$ (a constant derived from the Cassini state assumption for Mercury's rotation and playing the role of a mean obliquity), and the degree-two gravitational coefficients (\c20 and \d22), 
it is possible to determine the radius of the expected fluid core. 

Let A, B, and C be the principal moments of inertia of Mercury, $C_m$ the mantle's 
moment of inertia, $M$ the mass of Mercury, and $R$ its equatorial radius. The 
information on the core size is inferred from the following relation (Peale \etal 2002)~:
\begin{equation}
\left( \frac{C_{m}}{B-A} \right) \left( \frac{B-A}{MR^{2} }\right)  
\left( \frac{{MR^{2} }}{C} \right)  = \frac{C_{m}}{C} 
\leq 1
\label{eq:ratio}
\end{equation}
where a ratio of $C_{m}/C$ equal to 1 should indicate a core firmly coupled to the mantle 
and most likely solid. In the other hand, if the interior of Mercury is partly fluid, the ratio is smaller than 1.

The objective of the rotation parameter measurements consists in restraining each factor of the expression~(\ref{eq:ratio}). 
The first factor is inversely proportional to the $\varphi$ rotation angle (Peale \etal 2002). 
The second factor in~(\ref{eq:ratio}) is equal to 4\d22 while the third factor is related to 
$\bar{\eta}$ by the following relation (Wu \etal 1995)~:
\begin{equation}
\frac{MR^{2}}{C} = \frac{\mu}{n}\, \frac{ \sin{(I +
\bar{\eta})}}{\sin{\bar{\eta}}\, [ (1 + \cos{\bar{\eta}})\, G_{201}C_{22} -
(\cos{\bar{\eta}})\,G_{210}C_{20}]}
\label{eq:lawcassini}
\end{equation}

\vspace{3mm}
where $n$ is the mean orbital motion of Mercury and $\mu$ a constant of uniform precession of the orbit pole. $G_{201}$ and $G_{210}$ are the eccentricity functions of Kaula (Kaula, 1966). 
$G_{201} = \frac{7}{2}e - \frac{123}{16} e^{3}$ represent the first two terms in a series expansion where $e$ is the eccentricity while 
$G_{210} = (1-e^{2})^{-\frac{3}{2}}$. 
The formula~(\ref{eq:lawcassini}) expresses the Cassini state of Mercury (Colombo 1966; Peale 1969) where $I$ is the inclination of the orbit pole  of Mercury relative to the Laplace plane pole around which the orbit pole precesses at an uniform rate. Let us recall that in a Cassini state, the orbital and rotational parameters are indeed matched in such a way that the spin pole, the orbit pole, and the Laplacian pole remain coplanar while the spin and orbital poles on average precess at the same rate.

\vspace{0.5cm}
\centerline{Table 1 is here}
\vspace{0.5cm}

The rotational motion of Mercury is mainly characterized by a 3:2 spin-orbit resonance and a spin-orbit synchronism where axes of precession cones, spin and orbital, on average precess at the same rate. In the 2-body problem (Sun and Mercury), by using the 3:2 spin-orbit resonance and a few corollary approximations in the classic Euler equations of the 3-D rotational motion of Mercury distur\-bed by the solar torque, one may get two characteristic frequencies, namely the one in longitude $\Phi$, which 
is solution of the third Eulerian equation and the other in latitude of the body $\Psi$, which is solution of the two first Eulerian equations.
A Hamiltonian approach of the rotational motion of Mercury has been expanded by D'Hoedt $\&$ Lema\^{\i}tre (2004). The two resulting frequencies $\Phi$ and 
$\Psi$, usually called proper frequencies, calculated by this analytical approach and those obtained with our SONYR model are in a good agreement (see Table \ref{tab:Proper}). By the present paper, we have also improved the internal accuracy 
of our values (see Table \ref{tab:Proper}). 

\vspace{0.5cm}
\centerline{Table 2 is here}
\vspace{0.5cm}

Let us call $\eta$ the instantaneous obliquity of Mercury, defined as the angle bet\-ween Mercury's $Oz$ spin axis and the normal to its orbital plane; let $<\eta>$ be its mean value. Until the Radar observations recently obtained by Margot \etal 
(2006),\footnote{published at the very time when we are finishing the revised paper.}the value of Mercury's mean obliquity was not truly cons\-trai\-ned by measurements; many values were available in literature from zero to 7 arcminutes (amin hereafter). On the one hand, from the formula (2) computed with realistic Mercury's parameters, $\bar{\eta}$ 
can not be equal to zero. 
On the other hand, 7 amin given by Wu \etal (1995) seemed too high for Mercury. 
In our previous paper (Rambaux \& Bois 2004), using (2) we have estimated a range 
of $\bar{\eta}$ from 1.33 to 2.65 amin (for $I \in [5, 10]$ deg, 
$e \in [0.11, 0.24]$, with $\mu = {2 \pi}/\Pi$ and $n={2 \pi}/P_{\lambda}$ where 
values of the $\Pi$ orbital precession period and the $P_{\lambda}$ orbital period 
are given in Table \ref{tab:Periods}). 
However by using our SONYR model including the full spin-orbit motion of Mercury, for 
the plausible values of parameters we had chosen, the mean value of $\eta$ is close 
to 1.66 amin. In the present paper, by improving our method, we give a more relevant value for $<\eta>$ that proves to be in a very good agreement with the observations. 

We obtain the dynamical behavior of the hermean obliquity by the way of the following relation~:
\begin{equation}
\cos \eta = \cos i \cos\theta  + \sin i \sin\theta \cos(\Omega -\psi)
\label{eq:obl}
\end{equation}
where the spin-orbit variables are computed with SONYR. $i$ and $\Omega$ are respectively the inclination and the longitude of the ascending node of Mercury's orbital plane relative to the ecliptic plane. The angles $\psi$ and $\theta$ are the precession and nutation angles of the Eulerian sequence 3-1-3, while the third angle $\varphi$ is called the proper rotation. This angular sequence is used to describe the evolution of the body-fixed axis system $Oxyz$ (centered on the Mercury's center of mass) relative to a local reference system, dynamically non-rotating but locally transported with the translational motion of the body. Let $OX_1Y_1Z_1$ be the local reference system for the Mercury's rotation (see complete definitions of reference systems in Section 3).

In our previous work as well as in the present one, using our SONYR model, we have numerically integrated the complete equations of the \textit{N}-body problem (8 planets and the Sun) including the full spin-orbit motion of Mercury. More precisely, in our previous paper (Rambaux \& Bois 2004), we presented (i) an accurate theory of Mercury's spin-orbit motion, (ii) the proper frequencies related to the spin-orbit resonance, (iii) an analysis of main librations, and (iv) a first attempt for characterizing the dynamical behavior of Mercury's obliquity. The present paper is wholly devoted to stress and highlight the fourth point.

For lack of current accurate observations or ephemerides of Mercury's rotation, and therefore for lack of valid initial conditions for a numerical integration, we have built an original method for finding the libration center of the spin-orbit system and as a consequence for avoiding arbitrary amplitudes in librations of the spin-orbit motion as well as in Mercury's obliquity. Initial conditions of the rotational and orbital modes have indeed to be consistent between themselves with respect to a same spin-orbit reference system. As a consequence, using the SONYR model, the method requires in a first step to build mean initial conditions verifying geometrical conditions of a Cassini state for Mercury, that is to say a spin-orbit equilibrium state. In a second step, these mean initial conditions are fitted in order to place the spin-orbit system at its center of libration. The method has been carried out in two cases: (1) the spin-orbit motion of Mercury in the 2-body problem case (Sun-Mercury) where an uniform precession of the Keplerian orbital plane is kinematically added at a fixed inclination, (2) the spin-orbit motion of Mercury in the $N$-body problem (Sun and planets). 
With the spin-orbit system located at its center of libration in both cases above, the reference system of the first case is put at the average of perturbations of the second case. As a consequence, we make not only in evidence the strict dynamical behavior of Mercury's obliquity in both cases but may compare the respective oscillations. 

Our theoretical results are in a good agreement with recent radar observations.
However, it is not excluded that it should be possible to push farther the convergence process by drawing nearer still better to the libration center. In the present framework, we note that the dynamically driven spin precession, which occurs when the planetary interactions are included, is geome\-trically more complex than the purely kinematic case. In particular, we highlight that in the full spin-orbit motion of Mercury, the asynchronous rotation of the planet is really combined with a synchronism in precession variables as well as a second one between secular changes of the orbit inclination and the spin axis orientation.
As a consequence and supposing that it could exist an irreducible mini\-mum of the oscillation amplitude, we show how the quasi-periodic oscillations found in Mercury's obliquity could be geome\-trically understood. We discuss briefly the question to know whether the mi\-ni\-mal amplitude has to theoretically be of zero value or not. In the meantime of such a forthcoming work, we describe Mercury's rotation and the resulting equilibrium state that proceeds from a 2-body Cassini state. 

\section{The Spin-Orbit \textit{N}-bodY Relativistic model}

For obtaining the real motion of Mercury, we have used our model of solar System integration including the coupled spin-orbit motion of the Moon. This model (built by Bois, Journet, \& Vokrouhlick\'y 
and called BJV), expanded in a relativistic framework, had been previously built in accordance with the requirements of the Lunar Laser Ranging observational accuracy (Bois 2000; Bois \& Vokrouhlick\'y 1995). We extended the BJV model by generalizing the Moon's spin-orbit coupling to the other terrestrial planets (Mercury, Venus, the Earth, and Mars). The model is at present called SONYR (acronym of Spin-Orbit \textit{N}-BodY Relativistic model). As a consequence, the SONYR model gives an accurate simultaneous integration of the spin-orbit motion of Mercury. The approach of the BJV and SONYR models consists in integrating the \textit{N}-body problem (including translational and rotational motions) on the basis of the gravitation description given by the general relativity. The equations have been developed in the DSX formalism presented in a series of papers by Damour, Soffel, \& Xu
(1991, 1992, 1993). For purposes of celestial mechanics, to our knowledge, it is the most suitable formulation of the post-Newtonian (PN) theory of motion for a system of $N$ arbitrary extended, weakly self-gravitating, rotating and deformable bodies in mutual interactions. The DSX formalism, derived from the first PN approximation level, includes then the translational motions of the bodies as well as their rotational motions with respect to locally transported frames with the bodies. To each body is then associated a local reference system, centered on its center of mass and dynamically non-rotating. These local reference systems are freely falling in the local gravity field (the Sun) and are referred to a global reference system $O'X'Y'Z'$ for the complete {\it N}-body problem. 

A description of the SONYR model is given in our previous paper (Rambaux $\&$ Bois 2004). The principle of the model allows the analysis of Mercury's spin-orbit motion and the identification of internal mechanisms such as the direct and indirect planetary interactions. In addition, it gives also the interrelations between the parameters involved in Mercury's dynamical figure and the planet librations.

\section{Reference systems and their relations}

$O'X'Y'Z'$ is centered on the Solar System barycenter, the $O'X'Y'$ plane is parallel to the J2000 ecliptic plane. The $X'$-axis is oriented to the equinox $J2000$. Let $OXY$ 
be the mean orbital plane of Mercury and $OZ$ the mean orbital pole. Let us call 
$\mathcal{P}$ the instantaneous orbital plane of Mercury. The rotational motion of Mercury is evaluated from a body-fixed coordinate axis system $Oxyz$ centered on Mercury's center of mass relative to a local dynamically non-rotating reference system, $OX_1Y_1Z_1$. The $Ox, Oy, Oz$ axes are defined by the three principal axes of inertia $a, b$, and $c$ of the planet according to the best fitting ellipsoid such as $a > b > c$. The corresponding principal moments of inertia are then ordered as follows~: $A< B< C$. We used the Euler angles $\psi,\theta,\varphi$ related to the 3-1-3 angular sequence (3 represents a rotation around a $Z$-axis and 1, around an $X$-axis) to describe the evolution of the body-fixed axes $Oxyz$ with respect to axes of the local reference system $OX_1Y_1Z_1$. These angles are defined as follows~:
$\psi$ is the precession angle of the $Oz$ polar axis around the $OZ_1$ reference axis, $\theta$ is the nutation angle representing the inclination of $Oz$ with respect to 
$OZ_1$, and $\varphi$ is the rotation around $Oz$ and conventionally understood as the rotation of the largest rotational energy. This angle is generally called the proper rotation. The $Oz$ principal axis around which is applied the proper rotation is called the axis of figure and defines the North pole of the rotation (Bois 1995). The proper rotation $\varphi$ is around the smallest axis of inertia; it is called hereafter the spin angle and $Oz$ the spin axis. 

\vspace{0.5cm}
\centerline{Figure 1 is here}
\vspace{0.5cm}

\begin{figure}[!htb]
\begin{center}
\includegraphics[width=15cm]{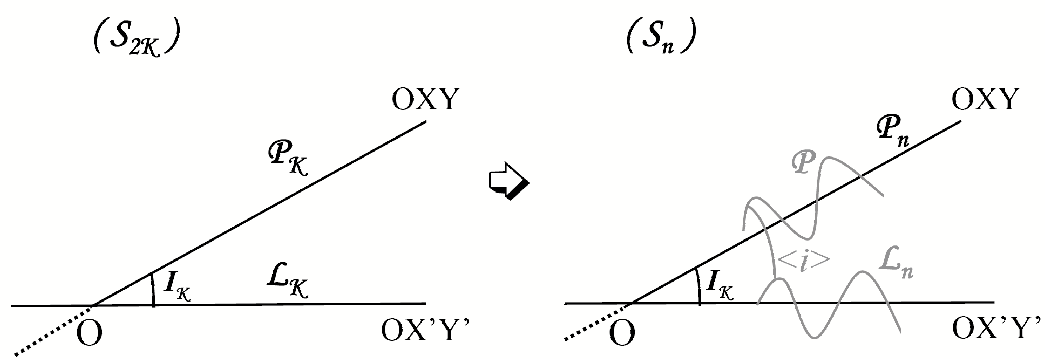}
\caption{Planes of reference of the $S_{2K}$ and $S_n$ cases linked by the conditions (4) (see Section 3).} 
\label{Fig1}
\end{center}
\end{figure}

In SONYR, building up the reference frames from the reference systems is related to the choice of the reference frames transported by the initial conditions (IC hereafter).  
This is of practical use when we have valid IC (for instance coming from ephemerides or derived from observations). In such conditions, the local reference frame for the rotation of a planet may be the one of the ephemeris. It is not actually the case for Mercury. True IC of Mercury's rotation are unknown (with respect to any reference system). Moreover, ad hoc attempts for determining instantaneous IC of the rotation produce a shift with those of the orbit and therefore a shift of the libration center of the spin-orbit resonance. Inconsistency between rotation's and orbit's IC drives necessarily to arbitrary amplitudes and phases in the spin-orbit librations. As a consequence, we have built (in Section 4) mean initial conditions (MIC hereafter) coming from geometrical conditions of the mean spin-orbit motion of Mercury. The latter is related to (i) the 3:2 spin-orbit resonance in cyclic variables ($\varphi$ and $\lambda$ such as 
$<d\varphi/dt>\;=(3n/2)\;+<d\omega/dt>$), (ii) the synchronism in precession variables 
($\psi$ and $\Omega$), and (iii) a second synchronism in nutation and inclination variables ($\theta$ and $i$) (see Rambaux \& Bois 2004). In a second step of the method (Section 5), the MIC are accurately corrected by taking into account a fit 
between the central axis of the precession cone described by the spin pole and the theoretical axis of reference. 
For these final MIC, the spin-orbit system is at its center of libration ($\Gamma$ hereafter) without arbitrary amplitudes. Moreover, we consider two cases~: 

(1) The spin-orbit motion of Mercury in the 2-body problem framework (Sun-Mercury) where a uniform precession, which we denote by $\Pi$, of the Keplerian orbital plane is kinematically added at a fixed inclination $I_K$ with respect to $OX'Y'$. Let us call 
$S_{2K}$ this construction bearing the geometrical Cassini state. Let us call 
$\mathcal{P}_K$ the associated orbital plane (a Keplerian plane in this case), and 
$\mathcal{L}_K$ the fixed plane of reference $OX'Y'$ ($\mathcal{L}_K$ plays the 
role of a Laplace plane for $\mathcal{P}_K$) (see Fig.~\ref{Fig1}).

(2) The spin-orbit motion of Mercury in the $N$-body problem framework (Sun and planets). Let us call $S_n$ this case. Let us call $\mathcal{P}_n$ the associated mean orbital plane of Mercury. The instantaneous orbital plane $\mathcal{P}$ has an instantaneous inclination $i$ relative to $OX'Y'$ while the inclination of $\mathcal{P}_n$ 
is $< i >$. Let us include the $S_{2K}$ construction into $S_n$ by putting 
$\mathcal{P}_K$ on $\mathcal{P}_n$. As a consequence, $I_{K}=\,< i >$. 
Over short periods, $\mathcal{P}_n$ remains equivalent to $\mathcal{P}_K$. 
The associated Laplace plane $\mathcal{L}_n$ is an instantaneous plane which 
follows the oscillations of $\mathcal{P}$ so that $i$ remains quasi-constant relative 
to $\mathcal{L}_n$. $\mathcal{L}_K$ co\"{\i}ncides with the average of the oscillations 
of $\mathcal{L}_n$, i.e. $<\mathcal{L}_n>\, \equiv \mathcal{L}_K$ (see Fig.~\ref{Fig1}). 

Our construction of reference systems verifies in the end the three following relations~:
\begin{equation}
\mathcal{P}_n \equiv \mathcal{P}_K   \;\;\;\; 
              < i >\,  \equiv I_K   \;\;\;\; 
              < \mathcal{L}_n >\,  \equiv  \mathcal{L}_K 
\end{equation}
which permit to accurately evaluate the obliquity's oscillations (i) not only with respect to the center of libration of the full spin-orbit motion of Mercury, but (ii) with respect to the included Cassini state. In other words, our construction is such as the mean Cassini state of the full spin-orbit motion of Mercury, on a given interval of time, co\"{\i}ncides with an exact Cassini state derived from the 2-body problem. 

Let us add that local dynamically non-rotating frames show in general relativity a slow (de Sitter) rotation with respect to the kinematically non-rotating frames. As a consequence, the $OX_1Y_1Z_1$ local reference frame for Mercury's rotation is affected by a slow precession of its axes transported with the translational motion of Mercury. In the Earth's case, the de Sitter secular precession of the Earth reference frame is close to 1.9"/cy (see Fukushima 1991; Bizouard \etal 1992; Bois \& Vokrouhlick\'y 1995). Consequently, the real rotation of Mercury has not to be expressed in an inertial system fixed in space, but in a local dynamically non-rotating frame fallen down in the gravitational field of the Sun. However, in the framework of the present paper where only Newtonian-like terms of SONYR are activated, $OX_1Y_1Z_1$ remains parallel to $OX'Y'Z'$.

\section{Building the mean initial conditions}

For lack of accurate observations or ephemerides of Mercury's rotation, and therefore for lack of valid IC for a numerical integration of the full spin-orbit motion of Mercury, a method is required for: (i) fitting the different parameters or constants involved in the spin-orbit of Mercury so that the SONYR model be self-consistent; (ii) removing arbitrary amplitudes in librations of Mercury's spin-orbit motion and therefore in its obliquity. In this section, we present the construction of MIC related to the spin-orbit coupling of Mercury. 

\vspace{2mm}
(1) We start from the geometry of the spin-orbit coupling problem defined in Goldreich \& Peale (1966): the spin axis of Mercury is considered normal to its orbital plane and the orbit is assumed to be fixed and unvarying. At this level, for expressing these assumptions, as well as that the long axis is pointed toward the Sun at a perihelion position $t_0$, analytical relations giving the corresponding set of initial conditions, namely $C_0$, can be written as follows~:
$$
(C_0) \left\{ \begin{array}{ccc}
\psi_0 = \Omega_0 & \;\;\;\;\;\;\; \theta_0 = I  \;\;\;\;\;\;\; & \varphi_0 = \omega_0 + \frac{3}{2}M_0 \\
\vspace{-5mm}
& &   \\   
\left ( \dfrac{d{\psi}}{dt} \right )_0= 0  &  \left (\dfrac{d \theta}{dt}\right)_0 = 0  &  
\left (\dfrac{d\varphi}{dt}\right)_0 =  \dfrac{3}{2}n  
\label{eq:C0i}
\end{array} 
\right. 
$$
\noindent where $\omega_0$ is the mean argument of periastron, and $M_0$ the mean ano\-ma\-ly with $\Omega_0=\omega_0=M_0=0$ at $t=t_0$. For such a Mercury's spin-orbit coupling based on a 2-body problem (with the asphericity factor 
$\alpha = \sqrt{3(B-A)/C}=0.0187$ and the eccentricity $e=0.206$), we have plotted a Poincar\'e section in Rambaux \& Bois (2004, Fig. 2), making in evidence the quasi-periodic librations around the 3:2 rotation state. 

\vspace{2mm}
(2) In the $S_{2K}$ framework, the condition of the precession synchronism drives to 
the geometrical property $<d\psi/dt>\,=2\pi/\Pi$ where $\Pi$ is the constant of uniform precession. The principle of our method requires to obtain the value of $\Pi$ from the $S_n$ case such as 
$2\pi/\Pi=\,<d\Omega/dt>$. In the same way, the value of $I=I_K$ is obtained from the $S_n$ case such as $I_K=\,< i >$, and as a consequence the value of $I$ is taken equal to $<i>$ since the $S_{2K}$ case. The mean initial conditions $C_{2K}$ for the $S_{2K}$ case are then written as follows~:
$$ 
(C_{2K}) \left\{ \begin{array}{ccc}
\psi_0 = \Omega_0 & \;\;\;\;\;\;\; \theta_0 = I = I_K  \;\;\;\;\;\;\; & \varphi_0 = \omega_0 + \frac{3}{2}M_0 \\
\vspace{-5mm}
& &   \\   
\left ( \dfrac{d{\psi}}{dt} \right )_0= \dfrac{2\pi}{\Pi} &  
\left (\dfrac{d \theta}{dt}\right)_0 = 0  &  
\left (\dfrac{d\varphi}{dt}\right)_0 = \dfrac{3}{2}n  
\label{eq:C2Ki}
\end{array} 
\right. 
$$
with $\Omega_0$, $\omega_0$, and $M_0$ equal to zero at $t=t_0$. 

\vspace{2mm}
(3) In the $S_n$ framework, the orbit precession of the rotating body is dynamically generated by the planetary interactions but its rate is nonuniform. However, $2\pi/\Pi$ has been rightly defined as $<d\Omega/dt>$. Moreover, the 3:2 spin-orbit resonance in the $S_n$ case involves the condition ${d\varphi}/{dt} = (2\pi/\Lambda) + (3n/2)$ where 
$\Lambda$ is obtained such as $2\pi/\Lambda=\,<d\omega/dt>$.
In addition, as shown further, $\theta$ and $i$ on average change at the same rate. This expresses a proximity to a Cassini state, which involves a third geometrical condition for building MIC, namely 
$<{d \theta}/{dt}>\,=\,<{di}/{dt}>\,= 2\pi/\Theta$. 
The mean initial conditions $C_{n}$ for the $S_{n}$ case are then written as follows~:
$$
(C_n) \left\{ \begin{array}{ccc}
\psi_0 = \Omega_0 & \;\;\;\;\;\;\; \theta_0 = \,< i >   \;\;\;\;\;\;\; & \varphi_0 = \omega_0 + \frac{3}{2}M_0 \\
\vspace{-5mm}
& &   \\   
\left ( \dfrac{d{\psi}}{dt} \right )_0= \dfrac{2\pi}{\Pi} &  
\left (\dfrac{d \theta}{dt}\right)_0 = \dfrac{2\pi}{\Theta} &  
\left (\dfrac{d\varphi}{dt}\right)_0 = \dfrac{2\pi}{\Lambda} + \dfrac{3}{2}n  
\label{eq:Cn}
\end{array} 
\right. 
$$
with $\Omega_0$, $\omega_0$, and $M_0$ equal to zero at $t=t_0$. 

\begin{figure}[!htb]
      \begin{center}
        \hspace{0cm}
        \includegraphics[width=8cm,angle=-90]{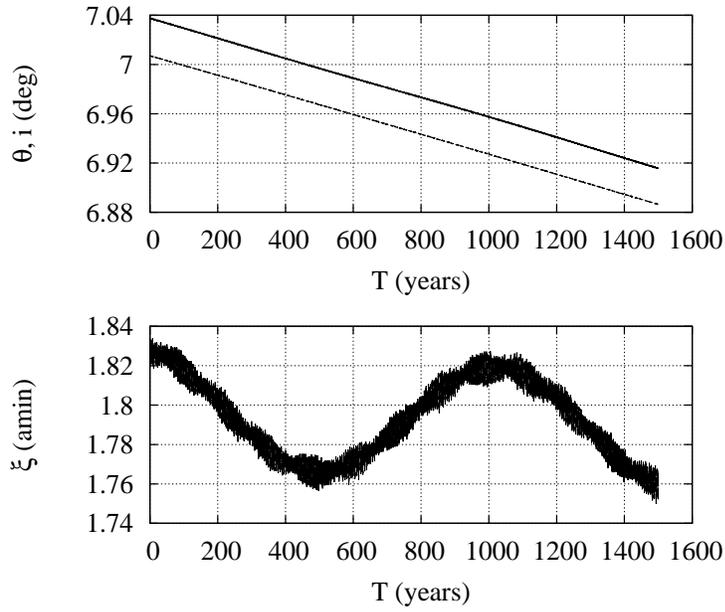}
        \caption{Second synchronism of the spin-orbit motion of Mercury. 
         $\theta$ and $i$ are plotted in the top panel over 1600 years.
         The difference $\xi = \theta - i$ plotted on the bottom panel librates
         within the proper frequency in latitude, namely $\Psi = 1066.8$ years. 
         Degrees (top panel) and arcminutes (bottom panel) are on the respective 
         vertical axes. Years are on both horizontal axes.}
        \label{Synchro2}
      \end{center}
\end{figure}

The dynamical behaviors of $\theta$ and $i$ have the same slopes 
(see  Fig.~\ref{Synchro2}, top panel). 
As a consequence, secular parts of $\theta$ and $i$ decrease at the same rate equal to $\Theta$. Although it is a question of a {\it N}-body problem, the spin keeps its proximity to a Cassini state during the slow change in $i$. But with this signature of the proximity to the Cassini state made in evidence in the $S_n$ framework, 
$<{d \theta}/{dt}>\,=\,<{di}/{dt}>$ may be understood as a second synchronism between the respective inclinations of the orbit and the spin axis relative to a same reference system ($OX'Y'Z'$). As a matter of fact we find that $\xi=\theta-i$ librates and rightly according to the $\Psi$ proper frequency in latitude (see  Fig.~\ref{Synchro2}, bottom panel).
 
\vspace{0.5cm}
\centerline{Figure 2 is here} 
\vspace{0.5cm}

\section{Fitting the mean initial conditions}

\begin{figure}[!htbp]
\begin{center}
\hspace{0cm}
\includegraphics[width=8cm,angle=-90]{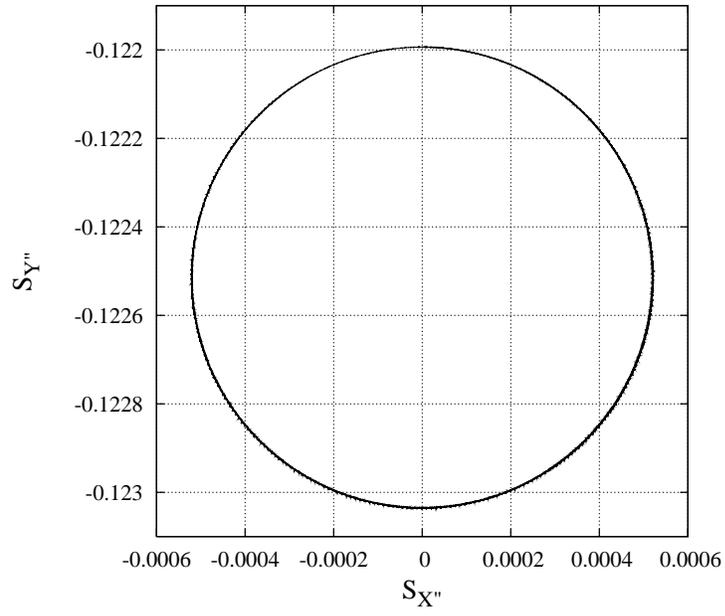}
\caption{Projection of Mercury's spin pole $S$ in mode SONYR($S_{2K}, C_{2K}$) on the $OX"Y"$ plane (i.e. rotating with $\Pi$).} 
\label{pole1}
\end{center}
\end{figure}

Let us call hereafter SONYR($S_i, C_i$) the integration of SONYR in the mode $S_i$ with the initial conditions $C_i$. Let us note that under the assumptions of the present paper, SONYR($S_i, C_i$) is a conservative dynamical system. Let $S$ be the $Oz$ spin pole of Mercury. $S$ describes a large cone with oscillations of polhodie; the projection of the cone in the $OX'Y'$ reference plane gives a large quasi-circle of center $A$ (with oscillations). Due to slightly heterogeneous initial spin-orbit positions, $A$ does not exactly co\"{\i}ncide with the projection of $OZ'$. By setting the latter in $A$, the system is placed at its $\Gamma$ libration center. This can be achieved by slightly moving $OX'Y'Z'$ or in SONYR by fitting a little bit the MIC. But because of large periods of precession, building and fitting the large circles is a very heavy computational process. Nevertheless, the reference frame that satisfies the whole system in $\Gamma$ is the one where all librations (related to small and large cones of precession) are at their respective right centers of libration. As a consequence, we have determined equations expressing the departure from $\Gamma$ within smaller circles of libration. Let us note $S_{X"}$ and $S_{Y"}$ the coordinates of $S$ projected onto a reference plane $OX"Y"$ rotating with the uniform motions $\Pi$ and $\Theta$; they are written as follows~:
\vspace{-1mm}
\begin{equation} 
\left\{ \begin{array}{ccc}
S_{X"} & = & \sin{[\psi - (\frac{2\pi}{\Pi}) t]} \sin{[\theta - (\frac{2\pi}{\Theta}) t]} \\  
S_{Y"} & = & -\cos{[\psi - (\frac{2\pi}{\Pi}) t]} \sin{[\theta - (\frac{2\pi}{\Theta}) t]} 
\label{eq:pole}
\end{array} 
\right. 
\end{equation}
where $2\pi/\Theta = 0$ in the $S_{2K}$ case. For each mode of SONYR ($S_{2K}$ and $S_{n}$), $S$ describes a small precession cone. Figure~\ref{pole1} gives the motion of $S$ in the mode 
SONYR($S_{2K}, C_{2K}$). Let us call $O\zeta$ the central axis of the small cone of precession whose coordinates are ($\zeta_1, \zeta_2$) in $OX"Y"$. 

\vspace{0.5cm}
\centerline{Figure 3 is here} 
\vspace{0.5cm}

\begin{figure}[!htbp]
\begin{center}
\hspace{0cm}
\includegraphics[width=8cm,angle=-90]{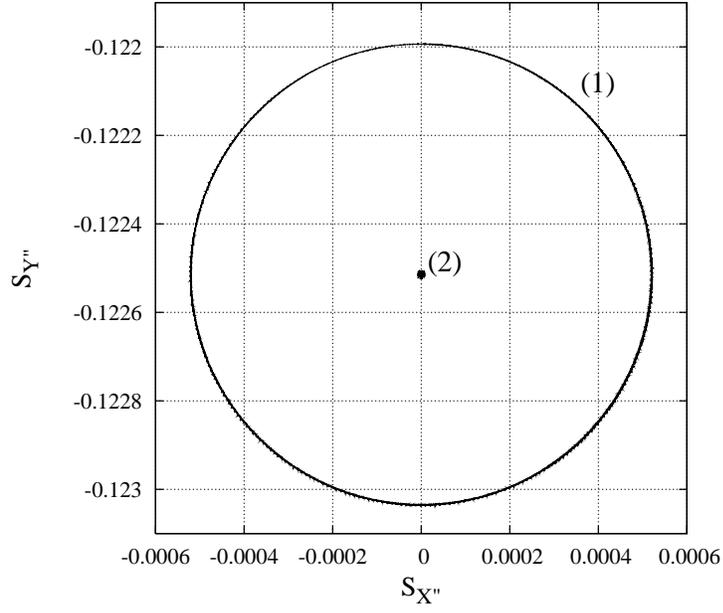}
\caption{Projection of Mercury's spin pole $S$ on the $OX"Y"$ plane. The large circle (1) obtained in mode SONYR($S_{2K}, C_{2K}$) is the one of Fig.~\ref{pole1}. In mode 
SONYR($S_{2K}, C'_{2K}$), it is a broadly reduced circle (2).} 
\label{pole2}
\end{center}
\end{figure}

The fitting equations are then written as follows~:
\vspace{-1mm}
\begin{equation} 
\left\{ \begin{array}{ccc}
S_{X"} & = & \zeta_1 \\
S_{Y"} & = & \zeta_2  
\end{array} 
\right. 
\label{zeta}
\end{equation}
After substitutions in orbital elements alone, the system (\ref{zeta}) becomes~:
\vspace{-1mm}
\begin{equation} 
\left\{ 
\begin{array}{ccc}
I_K + \delta I_K  & = & \arcsin{(\sqrt{\zeta_1^2 + \zeta_2^2 })} \\
\Omega_0 + \delta \Omega_0  & = & \arctan{(-\zeta_1/ \zeta_2) } 
\end{array} 
\right.
\end{equation}
where one gets $\delta I_K$ and $\delta \Omega_0$ as functions of  $I_K$, $ \Omega_0$, $\zeta_1$, and $\zeta_2$. In the mode SONYR($S_{2K}, C_{2K}$), we obtain then the two following  fitted initial conditions~:
\vspace{-1mm}
\begin{equation} 
\begin{array}{ccc}
\psi_0  & = & \Omega_0 + \delta \Omega_0  \\
\theta_0  & = & I_K + \delta I_K 
\end{array} 
\end{equation}
$C_{2K}$ becomes $C'_{2K}$ and one gets on the same graphic frame as previously a little circle (like a point at this scale, see Fig.~\ref{pole2}). The little circle expresses a considerable decreasing of the degree of opening of the cone by the above process.

\vspace{0.5cm}
\centerline{Figure 4 is here} 
\vspace{0.5cm}

\begin{figure}[!htbp]
\begin{center}
\hspace{0cm}
\includegraphics[width=8cm,angle=-90]{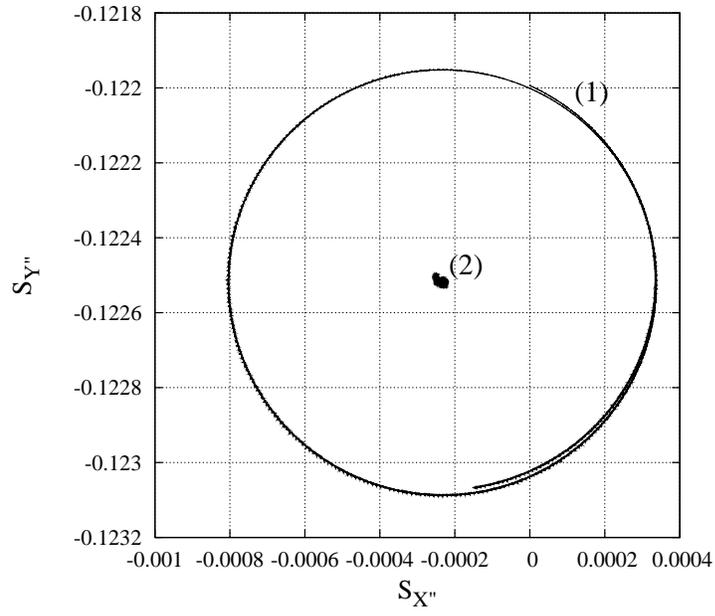}
\caption{Projection of Mercury's spin pole $S$ on the $OX"Y"$ plane. The large circle (1) 
is obtained in mode SONYR($S_{n}, C_{n}$) whereas the broadly reduced circle (2) in mode SONYR($S_{n}, C'_{n}$).} 
\label{pole3}
\end{center}
\end{figure}

The same principle is applied to the $S_n$ case with an analogous result. Taking into account in $C_n$ (becoming $C'_{n}$) the two fitted initial conditions $\psi_0$ and 
$\theta_0$ coming from (8) according to the same process, Figure~\ref{pole3} shows 
the significant decreasing of the degree of opening of the precession cone. 

\vspace{0.5cm}
\centerline{Figure 5 is here} 
\vspace{0.5cm}

\centerline{Table 3 is here} 
\vspace{0.5cm}

\section{The resulting obliquities}

\begin{figure}[!htbp]
\begin{center}
\hspace{0cm}
\includegraphics[width=8cm,angle=-90]{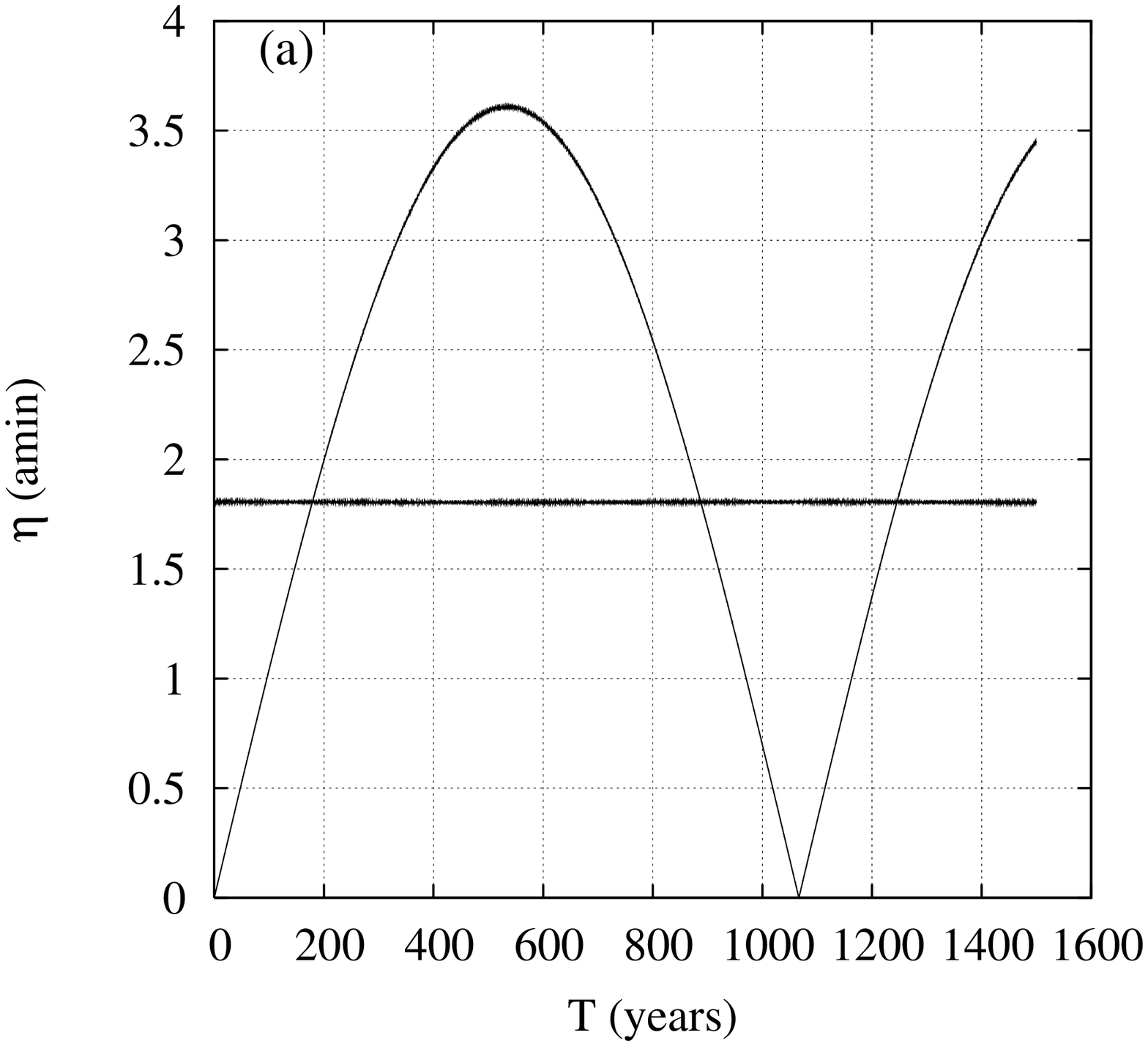}
\includegraphics[width=8cm,angle=-90]{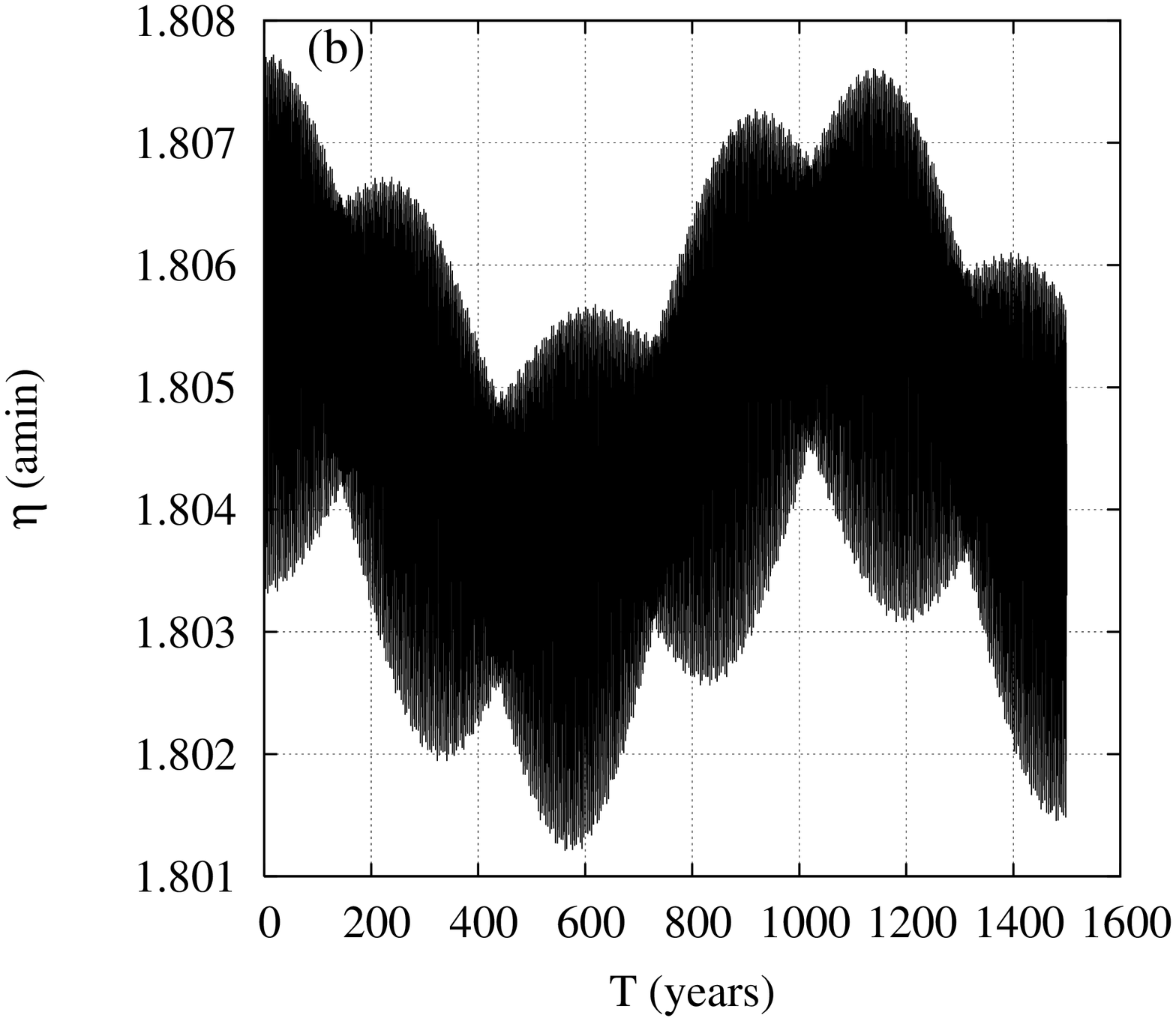}
\caption{(a) Resulting obliquities by SONYR($S_{2K}$, $C_{2K}$) and 
SONYR($S_{2K}$, $C'_{2K}$). (b) Zoom of $\eta$ in the SONYR($S_{2K}$, $C'_{2K}$) case, i.e. without arbitrary amplitudes. $\eta$ is characterized by a main oscillation according to the second proper frequency $\Psi$ within a peak-to-peak amplitude 
lesser than 0.4 as. In both panels, arcminutes are on the vertical axes and years 
on the horizontal ones.} 
\label{obl1}
\end{center}
\end{figure}

In the $S_{2K}$ mode, computations of $\eta$ give resulting obliquities before and after fitting the MIC at the libration center $\Gamma$, as shown in Figure~\ref{obl1}a. 
Figure~\ref{obl1}b shows a zoom of the obliquity purged of arbitrary amplitudes after 
the fitting process. However, $\eta$ appears not constant but quasi-constant. In the 
$S_{2K}$ mode indeed, i.e. without planetary interactions, due to the gravitational solar torque acting on the dynamical figure of Mercury with three unequal axes of inertia, 
the equations of the rotational motion solely form a non-integrable system whose 
regular solutions are quasi-periodic librations. As a consequence, the $\eta$ obliquity that depends on $\theta$ and $\psi$ (see formula (3) while $i=I_K$ and 
$\Omega=\Omega_0 + (2\pi/\Pi) t$ where $I_K$, $\Omega_0$ and $\Pi$ are constant values) follows in part this behavior. Nevertheless, with such faint amplitudes (peak-to-peak 0.4 arcseconds, 'as' hereafter), the obliquity may be reported quasi-constant in the $S_{2K}$ case, which is usually characteristic of a Cassini state derived from the 2-body problem (see Fig.~\ref{obl1}b). After removing the amplitude of the high frequency fluctuations (a modulation of a beat frequency with a period of 295 years whose peak-to-peak amplitude varies between a maximum near 0.28 as and a minimum near 0.14 as), the peak-to-peak amplitude of the resulting long period modulation of 1066.4 years is only 0.12 as.  

\vspace{0.5cm}
\centerline{Figure 6 is here} 
\vspace{0.5cm}

\begin{figure}[!htbp]
\begin{center}
\hspace{0cm}
\includegraphics[width=8cm,angle=-90]{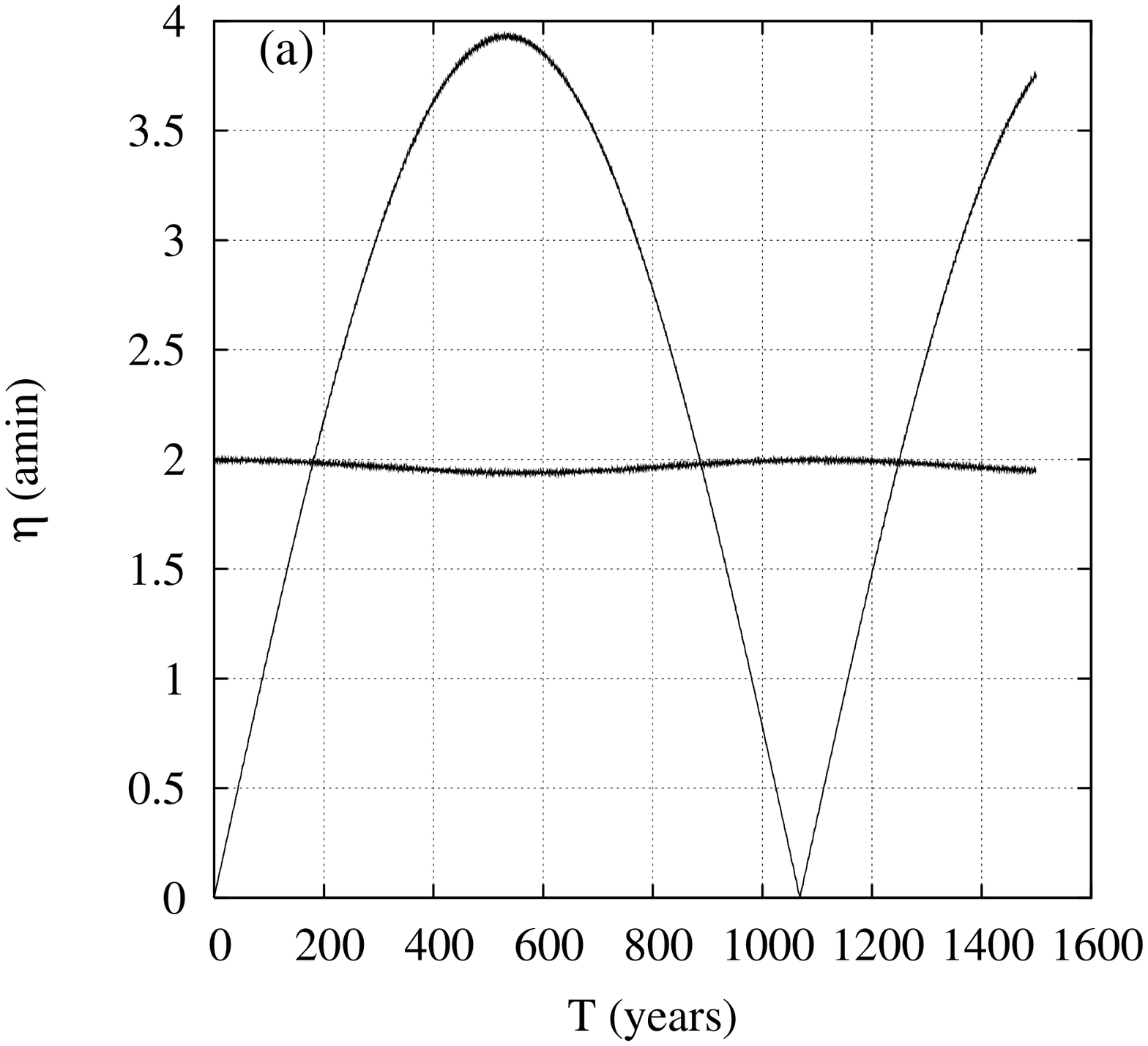}
\includegraphics[width=8cm,angle=-90]{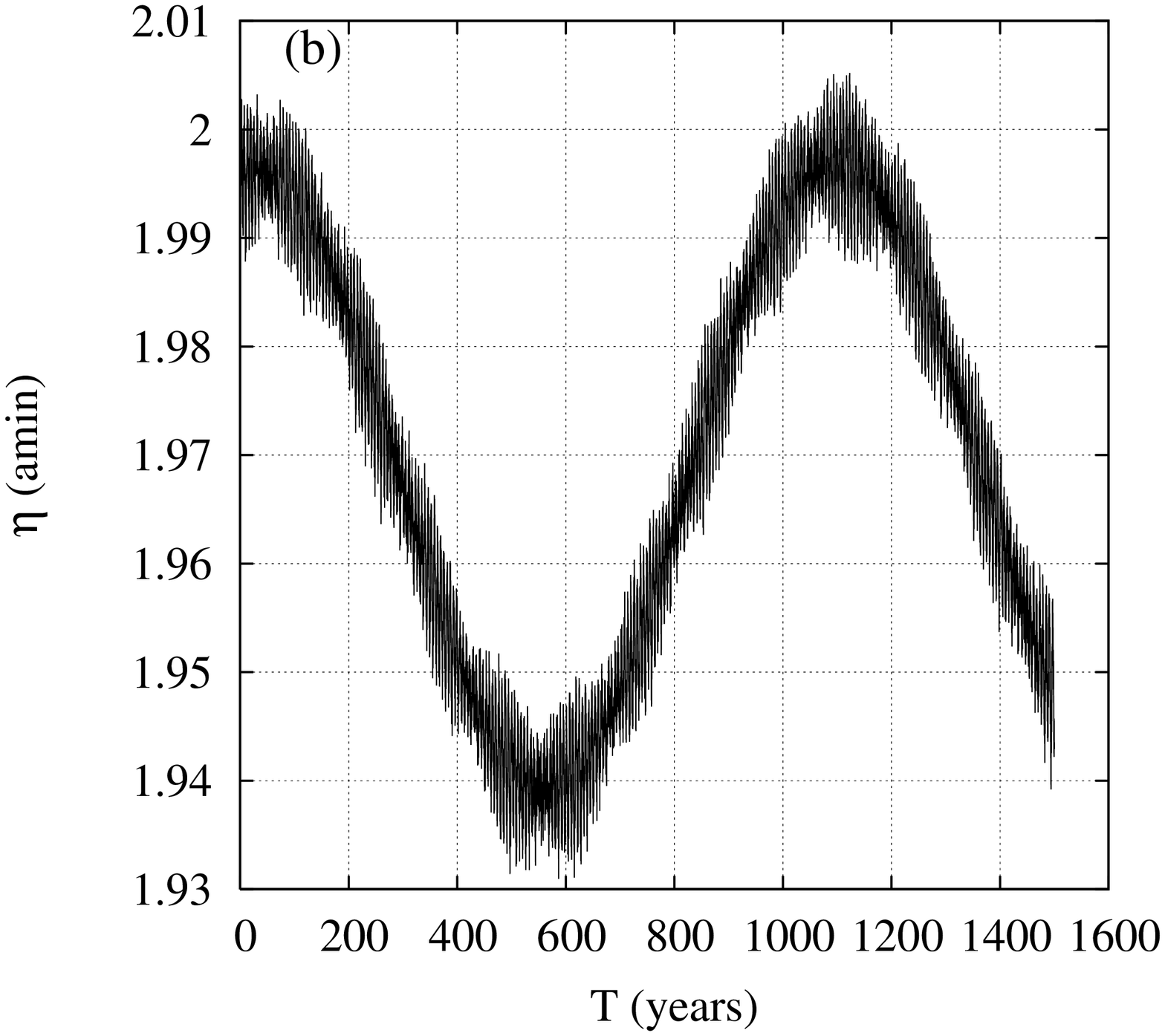}
\caption{(a) Resulting obliquities by SONYR($S_{n}$, $C_{n}$) and 
SONYR($S_{n}$, $C'_{n}$). (b) Zoom of $\eta$ in the SONYR($S_{n}$, $C'_{n}$) 
case, i.e. without arbitrary amplitudes. $\eta$ is charac\-terized by a clear oscillation according to the second proper frequency $\Psi$ while the peak-to-peak amplitude 
is near 4 as. Arcminutes are on the vertical axes and years on the horizontal ones.} 
\label{obl2}
\end{center}
\end{figure}

In the same way, in the $S_{n}$ mode, computations of $\eta$ give resulting obliquities, before and after the fitting process, as shown in Figure~\ref{obl2}a. Figure~\ref{obl2}b shows a zoom of the obliquity after the fitting process and then purged of arbitrary amplitudes. In both modes, $S_{2K}$ and $S_{n}$, the period of the main oscillation is 
$\Psi$, namely the second proper frequency of the resonance. A frequency analysis gives 1066.81 years in the $S_{n}$ case and 1066.44 years in the $S_{2K}$ one, let be 
a difference lesser than 0.4 years. In the $S_{n}$ case, after removing the amplitude 
of the high frequency fluctuations (about 0.6 as), the peak-to-peak amplitude of the resulting long period modulation of 1066.8 years is 3.4 as. 

\vspace{0.5cm}
\centerline{Figure 7 is here} 
\vspace{0.5cm}

Both sets of results, from $S_{2K}$ and $S_{n}$ cases are globally consistent with 
those of Peale (2006) and implicitly with those of Yseboodt \& Margot (2006). 
We note that the dynamically driven spin precession, which occurs when the planetary interactions are included, is geome\-trically more complex than the purely kinematic case. We find that the 3:2 spin-orbit resonance is not only combined with a first synchronism where the spin and orbit poles on average precess at the same rate but also with a second one where the rotational nutation desi\-gnated by $\theta$ 
and the $i$ orbit inclination on average decrease at the same rate (see 
Fig.~\ref{Synchro2} where is shown the libration of $\xi=\theta-i$ within the $\Psi$ period). These two synchronisms reflect the proximity to the Cassini state 
and their signatures express its conserving in the {\it N}-body problem case.
Besides, the conceptual question to know whether the ultimate minimal amplitude could be of zero value or not is discussed further. However, regardless of that, inside the present numerical process, the amplitudes in the $S_{n}$ case are related to a spin-orbit coupling. As shown below with the equation (9), it is about an indirect effect due to the solar torque, bearing the planetary interactions, and acting on the planet's rotation.

\begin{figure}[!htbp]
\begin{center}
\hspace{0cm}
\includegraphics[width=8cm,angle=-90]{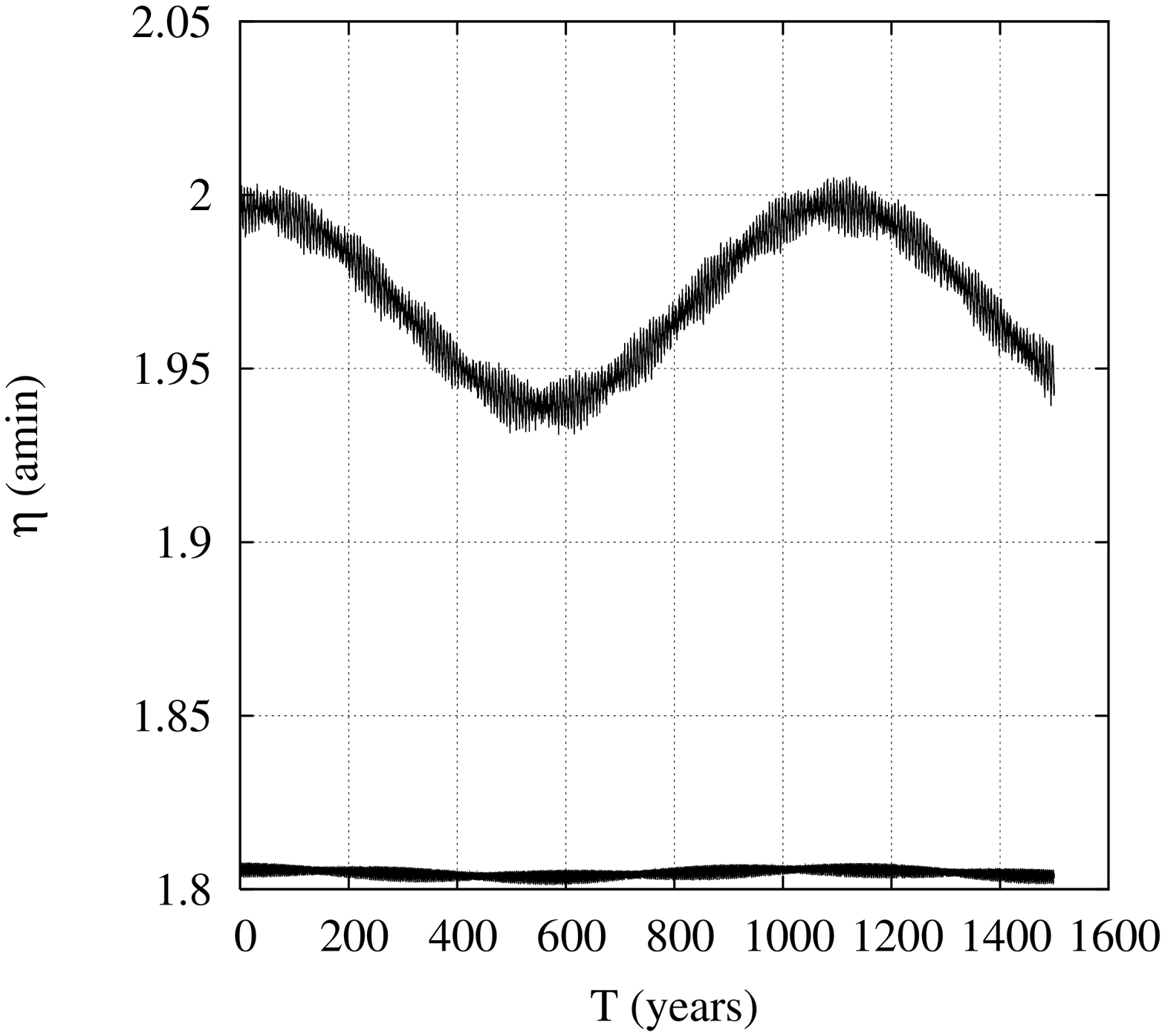}
\caption{Comparison of the obliquity's dynamical behavior in SONYR($S_{n}$, $C'_{n}$) and SONYR($S_{2K}$, $C'_{2K}$) modes, i.e. with and without planetary interactions, both computed and plotted at their respective libration center $\Gamma$. 
$<\eta>\,= 1.98$ amin in the $S_{n}$ case while 
$<\eta>\,= 1.80$ amin in the $S_{2K}$ case.} 
\label{obl3}
\end{center}
\end{figure}

Reference frames of the $S_{2K}$ and $S_{n}$ cases being linked by the conditions 
(4) (see Fig.~\ref{Fig1}) and values of the free constants required for 
SONYR($S_{2K}, C_{2K}$) being determined from SONYR($S_{n}, C_n$), 
SONYR($S_{2K}, C_{2K}$) has been built at the heart of SONYR($S_{n}, C_n$) 
in such a way the quasi-periodic solution of SONYR($S_{n}, C_n$) surrounds the vicinity of the quasi-periodic solution of SONYR($S_{2K}, C_{2K}$). Arbitrary amplitudes are not only avoided in the oscillations of $\eta$, but a direct comparison of $\eta$ coming from $S_n$ with $\eta$ coming from $S_{2K}$ is possible, which makes in evidence the effect of the planetary interac\-tions. In the $S_{n}$ mode indeed, the gravitational solar torque, bearing the planetary interactions and acting on the dynamical figure of Mercury, induces more sizeable librations related to the disturbed Euler angles as well as the variables 
$i$ and $\Omega$. At this phase of analysis, the obliquity can be no longer said 
quasi-constant. The oscillation amplitude of period $\Psi$ is indeed equal to 4 as 
(total peak-to-peak amplitude), let be one order of magnitude larger than in the 
$S_{2K}$ case (or 28 times larger when comparing the peak-to-peak amplitudes 
of the strict $\Psi$ modulation without high frequency fluctuations) 
(see Fig.~\ref{obl1}b and Fig.~\ref{obl2}b).
The comparison of the $S_{2K}$ and $S_n$ modes, respectively derived from the 
$2$-body and $N$-body problems, both computed and plotted at their respective libration centers $\Gamma$, shows how Mercury's obliquity and its dynamical behavior are modified when the planetary interactions are fully introduced (see Fig.~\ref{obl3}). The mean obliquity is larger as well as the oscillation amplitude. $<\eta>\,= 1.98$ amin 
in the $S_{n}$ case while $<\eta>\,= 1.80$ amin in the $S_{2K}$ case. 

This spin-orbit mechanism is rightly an indirect effect due to the solar torque, bearing the orbital planetary interactions and acting on Mercury's rotation, as shown by the classic general equation of the rotational motion of a solid celestial body (Bois 1995)~: 

\begin{equation}
    \frac{\partial [(I) \pmb{ \omega}] }{\partial t} + 
    \pmb{ \omega} \times [(I) \pmb{\omega} ] = \left(\frac{3Gm}{a^{3}}\right)\left(\frac{a}{r}\right)^{3}
    \pmb{ u} \times (I) \, \pmb{ u}
    \label{eq:Liouville}
\end{equation}

where $\pmb{\omega}$ is the instantaneous rotational vector of Mercury and $(I)$ its tensor of inertia. The right hand side of the equation represents the second degree torque exerted by a point mass $m$ (the Sun) of position $r\pmb{ u}$ with respect to the body (Mercury) when it is reduced to three oblateness coefficients. $a$ is the mean distance between the two bodies while $r$ is the instantaneous one. $r\pmb{ u}$ is solution of the $N$-body problem in SONYR($S_{n}$). Let us note that the direct planetary torques are about two orders of magnitude smaller than the indirect effects acting by the solar torque. 

\vspace{0.5cm}
\centerline{Figure 8 is here} 
\vspace{0.5cm}

Due to the planetary interactions, Mercury's orbit is no longer Keplerian and rate of precession of Mercury's orbital plane is no longer uniform. As a consequence, the obliquity's dynamical behavior is no longer quasi-constant. More generally, it is not possible for the obliquities to be constant since the orbit precession occurs at non-uniform rates (Bills 2005). In the case of Mercury, according to Bills \& Comstock (2005), if the orbit pole precession rate were uniform, dissipation would have driven Mercury into a Cassini state, in which the spin pole and orbit pole remain coplanar with the Laplace pole, as the spin pole precesses about the moving orbit pole. However, according to these authors and in accordance with our results, due to the nonuniform orbit precession rate with several distinct modes of oscillation, this simple coplanar configuration is no longer globally maintained. 

\section{Discussion}

We have significantly reduced amplitudes of the oscillations in Mercury's obli\-quity, which were exposed to discussion since a previous paper (Rambaux \& Bois 2004). 
For doing that, we have built an original method for finding the libration center of the spin-orbit system and as a consequence for avoiding arbitrary amplitudes in librations 
of the spin-orbit motion as well as in Mercury's obliquity. We have converged and the amplitudes are in very good agreement with most recent observations (Radar observations by Margot \etal 2006). This validates not only our theoretical SONYR 
model applied to Mercury but also our method of convergence.

Our centering method has converged but it is not excluded that it may remain a residual amplitude of arbitrary oscillations. It ought to be possible to push farther the convergence process by drawing nearer yet more precisely to the libration center (in 
the framework of the paper, i.e. in a conservative dynamical system). Since our amplitudes are compatible with current observations, we have concluded that this question could wait for MESSENGER and BepiColombo data in order to directly fit 
the SONYR model to observations.

Nevertheless, apart from values of residual amplitudes, it remains a conceptual question. Does there exist a theoretical possibility to reach a pure point of equilibrium where the oscillation amplitude would be absolutely of zero value or, on the contrary, does there exist an irreducible minimum, even very small but of non-zero value?
In the $S_{n}$ case, existence of distant resonances of high orders that would excite 
the 3:2 resonance is not excluded a priori. None mathematical theorem does deny it 
but the demonstration of this would take us along to a purely mathematical ground. Consequently, we will say that the question remains open (both in conservative and dissipative cases). However, we plane some further works in this connection. The occurrence of a perfect or rigid equilibrium without oscillations seems to us unlikely 
for natural systems.

\section{Conclusion}

Using our SONYR model of the solar System including the spin-orbit couplings of Earth-like planets, and setting the system at its libration center in order to eliminate arbitrary amplitudes (at the internal accuracy level of the model), we have obtained the complete obliquity for the full spin-orbit motion of Mercury. 
Our construction of reference systems permits to accurately evaluate the obliquity's oscillations (i) not only with respect to the libration center of the full spin-orbit motion 
of Mercury ($S_n$), (ii) but also with respect to a Cassini state derived from the 2-body problem ($S_{2K}$). The construction is such as the exact Cassini state derived from 
$S_{2K}$ co\"{\i}ncides with the mean Cassini state of $S_n$ on a given interval of time. 
We have obtained (1) a mean value of Mercury's obliquity that is in a very good agreement with very recent observations (Radar observations by Margot \etal  2006), namely 1.98 amin with SONYR for $2.1 \pm 0.1$ amin by observations, (2) amplitudes of oscillations in Mercury's obliquity of the order of 2 as, which is compatible with the results of these observations where the deviation with respect to the 2-body Cassini state is constrained lesser than 6 as. These observational data validate not only our theoretical prediction with the SONYR model applied to Mercury but also our method of convergence for finding a libration center in a complex spin-orbit system. 

Being dependent on the question of existence of an irreducible minimum in the oscillation amplitude, we find that the equilibrium state of Mercury computed in the framework of a conservative $N$-body system squares to a quasi-periodic oscillation without secular drifts. In other words, in spite of the planetary interactions indirectly acting by the solar torque on Mercury's rotation, Mercury remains in a stable equilibrium state that proceeds from a 2-body Cassini state. 

Besides, by the way of our method, we have made in evidence the richness of a dynamical precession related to planetary interactions ($N$-body problem) relative to a purely kinematical orbit precession added to a 2-body problem (Sun and Mercury). 
Whatever the open question on the minimal amplitude in the obliquity's oscillations, the approach of the 2-body Cassini state is necessarily limited to reach a constant or quasi-constant behavior of Mercury's obliquity (where moreover $<\eta>\,= 1.80$ amin versus $1.98$ amin, for a difference of $10.8$ as). The planetary interactions generate a forced dynamical precession of Mercury's orbital plane with a nonuniform rate. This precession is kinematically taken into account in the conventional construction of a Mercury's Cassini state as well as in our $S_{2K}$ case. But in the $S_n$ case, due to the first synchronism where the spin and orbit poles on average precess at the same rate (as well as the second one where the spin axis orientation, namely the $\theta$ nutation, and the orbit inclination $i$ on average change at the same rate), the difference angle between these two poles, i.e. $\eta$, (and $\theta-i$ as well) librates necessarily. 
As a consequence and whether it would turn out that there exists an irreducible minimum of the oscillation amplitude, so these resulting quasi-periodic oscillations in the obliquity ought to be understood as librations related to these two synchronisms where the 
$(\psi-\Omega)$ and $(\theta-i)$ respective differences librate (according to the $\Psi$ 
proper frequency in latitude).
In the same way, in multi-planet systems, when the longitudes of periapse on average precess at the same rate, the difference angle between the two apsidal lines librates 
(see e.g. Lee \& Peale 2003 or Bois \etal 2003). 
Under this assumption and according to the obtained dynamical behavior, the obliquity 
of Mercury should be characterized by a quasi-periodic oscillation according to the 
period $\Psi$, with an amplitude $A_{\eta}$ and a mean obliquity $<\eta>$, in such a 
way that Mercury's obliquity $\eta$ could be written as follows (in first approximation)~:
\begin{equation}
\eta =\,<   \eta >  + A_{\eta}\cos{\left(\dfrac{2\pi}{\Psi} t + \alpha\right)}
\end{equation}
\noindent where $\alpha$ is a phase factor.

In the end, for lack of current and accurate observations or ephemerides of Mercury's rotation, but with SONYR($S_{n}, C'_n$) set up at the libration center of Mercury's spin-orbit motion, we may conclude that our method is relevant to link physical causes and their respective effects. As a consequence, with accurate observations of Mercury's rotational motion, that should be being achieved by MESSENGER and BepiColombo, SONYR($S_{n}, C'_n$), and not a 2-body Cassini state model as 
SONYR($S_{2K}, C'_{2K}$), ought to be fitted to these observations.

\vspace{3mm}
\textit{Acknowledgments}
The authors thank the anonymous referee and Bruce Bills for their constructive requirements. NR acknowledges the ROB and FUNDP for their fellowship supports.

\newpage

\vskip -0.4truecm
\begin{table}[!htb]
\centering 
\begin{tabular}{lcc}
\\
\hline
\hline
Proper Frequencies & $\ \ $Ê$\Phi$ (years) $\ \ $ & $\ \ $ $\Psi$ (years) $\ \ $ \\
\hline 
Rambaux $\&$ Bois 2004 & 15.847 & 1066 \\
D'Hoedt $\&$ Lema\^{\i}tre 2004 & 15.857  &  1066.68 \\
This paper & 15.824 $\pm$ 0.024 & 1066.8 $\pm$ 0.4  \\
\hline
\end{tabular}
\caption{Proper frequencies related to the spin-orbit motion of Mercury
 (in longitude: $\Phi$, in latitude: $\Psi$). Comparaison of results.}
\label{tab:Proper}
\end{table}

\newpage 

\vskip -0.4truecm
\begin{table*}[!htb]
\centering     
 \begin{tabular}{lcr}
 \\
\hline
\hline
Basic periods \\
\hline
$P_{\varphi}$ (rotation period) & =  & 58.646 days \\
$P_{\lambda}$ (orbital period) &  = & 87.969 days \\
$\Pi$ (orbital precession) & = & 279 000 years \\
\hline        
\end{tabular}
\caption{Basic periods of the spin-orbit motion of Mercury (Rambaux \& Bois 2004)}
\label{tab:Periods}
\end{table*} 

\newpage

\begin{table}[htdp]
\begin{center}
\begin{tabular}{l|ll}
& $S_{2K}$ & $S_{n}$ \\
\hline
\hline
$\delta \Omega_0$ (amin) & 0.0   & -6.564 \\
$\delta I_K$ (amin) & -1.806  & -1.823
\end{tabular}
\caption{Numerical values of $\delta \Omega_0$ and $\delta I_K$ in the  
$S_{2K}$ and $S_{n}$ cases.}
\end{center}
\label{default}
\end{table}

\newpage

\vspace{3cm}

\newpage 

\newpage 

\newpage 

\newpage 

\newpage 

\newpage 

\newpage 

\begin{thebibliography}{}

\bibitem{B.G.:2000gq} Bills, B.G., 2005, Free and forced obliquities of the Galilean satellites of Jupiter, Icarus 175, 233

\bibitem{Bills:2005my} Bills, B.G. \& Comstock, R.L., 2005, Forced obliquity variations of Mercury, Journal of Geophysical research 110, 1

\bibitem{Biz} Bizouard, C., Schastok, J., Soffel, M., 
Souchay, J., 1992, in~: N. Capitaine (ed.), Journ\'ees 1992~: Syst\`emes 
de r\'ef\'erence spatio-temporels, Observatoire de Paris, 76

\bibitem{Bois95} Bois, E., 1995, Proposed Terminology for a General Classification of Rotational Swing Motions of the Celestial Solid Bodies, A\&A 296, 850

\bibitem{Bois00} Bois, E., 2000, Knowledge of the lunar librations at the Lunar Laser Ranging experiment epoch, C. R. Acad. Sci. Paris, t. 1, S\'erie IV, 809

\bibitem{BoisVok95} Bois, E., \& Vokrouhlick\'y, D., 1995, Relativistic spin effects in the Earth-Moon system, A\&A 300, 559

\bibitem{Bois03} Bois, E., Kiseleva-Eggleton, L., Rambaux, N., \& Pilat-Lohinger, E., 2003, Conditions of Dynamical Stability for the HD 160691 Planetary System,  Astrophysical Journal 598, 1312

\bibitem{lee03} Lee, M.H., \& Peale, S., 2003, Secular Evolution of Hierarchical Planetary Systems., 2003, Astrophysical Journal 592, 1201 

\bibitem{Colombo3} Colombo, G., 1965, Rotational period of the planet Mercury, Nature 208, 575

\bibitem{Colombo} Colombo, G., 1966,  Cassini's second and third laws, AJ 71, 891

\bibitem{Correia} Correia, A., \& Laskar, J., 2004, Mercury's capture into the 3/2 spin-orbit resonance as a result of its chaotic dynamics, Nature 429, 848

\bibitem{Damour91}
{Damour}, T., {Soffel}, M., \& {Xu}, Ch. 1991, 
\newblock  General-relativistic celestial mechanics. I. Method and definition of reference systems, Phys. Rev. D 43, N. 10, 3273

\bibitem{Damour92}
{Damour}, T., {Soffel}, M., \& {Xu}, Ch. 1992, 
\newblock  General-relativistic celestial mechanics. II. Translational equations of motion,
Phys. Rev. D 45, N. 4, 1017

\bibitem{Damour93}
{Damour}, T., {Soffel}, M., \& {Xu}, Ch. 1993, 
\newblock  General-relativistic celestial mechanics. III. Rotational equations of motion,
Phys. Rev. D 47, N. 8, 3124

\bibitem{sandrineanne} D'Hoedt, S., \& Lema\^{\i}tre, A., 2004, The spin-orbit resonant rotation of Mercury: a two degree of freedom Hamiltonian model, Celest. Mech. and Dyn. Astron. 89, 267

\bibitem{Fuk1991} Fukushima, T., 1991, Geodesic nutation, A\&A 244, L11

\bibitem{gold2}Goldreich, P., \& Peale, S., 1966, Spin-orbit coupling in the solar system, AJ 71, N. 6, 425

\bibitem{Kaula} Kaula, W.M., 1966, Theory of Satellite Geodesy; Applications of Satellites to Geodesy, Blaisdell, Waltham, MA

\bibitem{Laskar} Laskar, J., \& Robutel, P., 1993, The chaotic obliquity of the planets, Nature 361, 608

\bibitem{Margot} Margot, J.-L., Peale, S.J., Slade, M.A., Jurgens, R.F., \& Holin, I., 
2006, Obervational proof that Mercury occupies a Cassini state, AAS, DPS Meeting \#38, \#49.05

\bibitem{Milani} Milani, A., Vokrouhlick\'y, D., \& Bonanno, C., 2001, Gravity field and rotation state of Mercury from the BepiColombo Radio Science Experiments, Planet. Space Sci. 49, 1579

\bibitem{Moons82} 
{Moons}, M., 1982,
\newblock Analytical theory of the libration of the moon, 
\newblock  Moon and the Planets 27, 257

\bibitem{Peale1} Peale, S.J., 1969, Generalized cassini's laws, AJ 74, 483

\bibitem{Peale2} Peale, S.J., 1972, Determination of parameters related to the interior of Mercury, Icarus 17, 168

\bibitem{Peale3} Peale, S.J., 2006, The proximity of Mercury's spin to Cassini state 1 from adiabatic invariance, Icarus 181, 338

\bibitem{Peale04} Peale, S.J., Phillips, R.J., Solomon, 
S.C., Smith, D.E., \& Zuber, M.T., 2002, A procedure for determining the nature of Mercury's core, Meteo. and Planet. Sci. 37, 1269

\bibitem{Pett} Pettengill, G.H., \& Dyce R.B., 1965, A Radar determination of the rotation of the planet Mercury, Nature 206, 1240

\bibitem{RambauxBois03a}
{Rambaux}, N., \& {Bois}, E., 2004, Theory of the Mercury's spin-orbit motion and analysis of its main librations, A\&A 413, 381
        
\bibitem{sol} Solomon, S.C., 20 colleagues, 2001, The MESSENGER mission to Mercury: scientific objectives and implementation, 
Planet. Space Sci. 49, 1445

\bibitem{Wu95} Wu, X., Bender, P.L., \& Rosborough, G.W., 1995, Probing the interior structure of Mercury from an orbiter plus single lander, Planet. Space Sci. 45, 15

\bibitem{Yse06} Yseboodt, M., \& Margot, J.L., 2006, Evolution of Mercury's obliquity, Icarus 181, 327

\end{thebibliography}
\end{document}